\documentclass{emulateapj}
\usepackage{amssymb,amsmath}
\usepackage{natbib}
\usepackage{graphicx}
\usepackage{color}
\usepackage{sidecap}
\usepackage{float}

\definecolor{Forest}{rgb}{0,0.5,0.10}
\definecolor{Gray}{rgb}{0.6,0.6,0.6}

\begin{document}
\title{
Sterile and Fertile Planetary Systems \\
Statistical Analysis of Multi-Planet Systems in Kepler's data
}
\author{Amir Weissbein\altaffilmark{1}, Elad Steinberg\altaffilmark{1} and Re'em Sari\altaffilmark{1,2}}
\altaffiltext{1}{Racah Institute of Physics, Hebrew University, Jerusalem 91904, Israel}
\altaffiltext{2}{Theoretical Astrophysics, Caltech 350-17, Pasadena, CA 91125, USA}

\email{amir.weissbein@mail.huji.ac.il, elad.steinberg@mail.huji.ac.il}
%\maketitle

\begin{abstract}
The Kepler mission has discovered a large number of planetary systems.
We analyze the implications of the discovered single/multi-exoplanet systems from Kepler's data.

As done in previous works,
we test a simple model in which the intrinsic occurrence of planet is an independent process, and with equal probability around all planet producing stars. This leads to a Poisson distribution for the intrinsic number of planets around each host.
However, the possibility of zero/low mutual inclination is taken into account, creating a correlation between detecting different planets in a given stellar system, leading to a non Poisson distribution for the number of transiting planets per system.
Comparing the model's predictions with the observations made by Kepler, we find that the correlation produced by planarity is insufficient and a higher correlation is needed;
either the formation of one planet in the system enhances the likelihood of other planets to form, and/or that some stars are considerably more fertile than others.
Kepler's data presents evidences that both correlations might play a part, in particular a significant dependency in the radial distribution of planets in multi-planet systems is shown.
Followup observations on Kepler planet's hosts can help pinpoint the physical nature of this correlation.
\end{abstract}

\section{Introduction}

The Kepler satellite \cite[]{kep_sci} has opened a new era in exoplanet research.
Since Kepler's launch over 2300 exoplanet candidates
have been reported, most of which are Neptune sized or smaller \cite[]{K2011,K2012}.
In the following, we refer to planet candidates as planets due to the low occurrence of false positives \cite[]{MJ2011,L2012}.
The observational data from Kepler's first four and sixteen months of activity respectively as presented in \cite{K2011} and \cite{K2012},
revealed a large number of multi-exoplanet systems.
The large amount of data gives a significant boost to the statistical discussion on multi-exoplanet systems.
In particular, one may deduce from the data information about the mutual influences of planets on each other \cite[]{RH2010,Lis2011,TD2011}.

The simplest model, which assumes that the occurrence of transiting planets is an independent process and that all stars are equally likely to produce planets, results in a Poisson distribution of the number of
transiting
planets in a system.
However, the observed transiting planet distribution from Kepler has been shown not to fit a Poisson distribution \citep{Lis2011}.
This can easily be seen by comparing the ratio of the number of systems with one detected planet to the number of systems with two detected planets from the data published in \cite{K2012}. For a Poisson distribution, this ratio is given by $\mu/2$, where $\mu$ is the mean of the Poisson distribution.
Therefore, we expect to have approximately $\mu=0.344$.
Using this mean, it follows that there should only be 2.4 systems with four transiting planets, 0.16 with five transiting planets and $9\cdot 10^{-3}$ systems with six transiting planets.
These numbers are very different from the observed data which has 27 systems with four transiting planets, 8 systems with five transiting planets and one system with six transiting planets (see \S\ref{od}).
It is evident that the real distribution has a longer tail since Kepler's data indicates that there are more systems with a large number of transiting planets than predicted from a Poisson distribution.

The long tail can arise if the planet occurrence is a dependent process, where the formation or detection of one planet enhances the probability for another planet to form, and/or if solar systems tend to be rather planar \citep[e.g.,][]{Lis2011}.
In this paper we examine the predictions of a planar and low mutual inclination planetary systems under the assumption that planet occurrence is independent.
In such planar and low mutual inclination independent models, while the true occupation number is Poisson, the transiting planet distribution is not, since planarity implies that if a given planet transits, others are more likely to transit as well.
We compare the predictions to Kepler's data. We see that while planarity indeed creates a longer
tail than in a Poisson distribution, it is not yet enough to explain the planet number distribution.
The structure of the paper is as follows:
In \S \ref{planar} we present the analytical planar independent model and compare it to Kepler's observations in \S \ref{obs_data}.
We then expand the planar independent model in \S \ref{nonplanar} to include finite mutual inclinations in the planetary systems. We show that neither
can explain Kepler's data and some correlation is necessary. In \S \ref{sec:depend} we address the possible reason for the model's failure and in \S\ref{sec:compare} we compare the independent model to previous works.
Finally in \S\ref{summary} we discuss possible origins for these correlations.

\section{Kepler's Data and The Intrinsic Planetary Distribution}\label{od}
Following \cite{H2011} and \cite{TD2011}, we restrict Kepler's detections to the well determined F, G and K
stars.
Like \cite{TD2011}, we only investigate stars that have $\log g\geq 4\ $, $\ 4000^oK\leq T_{eff}\leq 6500^oK\ $ and Kepler magnitude $9\leq K_{p}\leq 16$, where the data was taken from Kepler's published result in \cite{KIC,K2012}
\footnote{http://archive.stsci.edu/kepler/planet\textunderscore candidates.html}.

\begin{figure}
  % Requires \usepackage{graphicx}
 % \includegraphics[width=0.8\textwidth]{distribution2.eps}\\
 \plotone{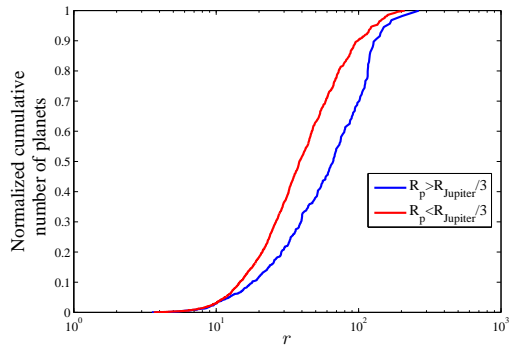}
  \caption{The normalized cumulative number of planets as a function of $r$ for small and big planets.
The black dashed line denotes $r=75$, the semi major axis in which the small planets experience a role over
}\label{fig:cumulative}
\end{figure}

Since completeness of the sample is required for a proper statistical analysis,
we must restrict the discussion to planets with short enough periods.
This restriction insures that the data can be assumed to be complete, i.e.
further data releases will not change significantly the observed data up to the cut-off.
In order to evaluate, $r_{\max}$ ,the upper boundary of our sample, we calculate the cumulative number of planets which satisfies the above criteria as a function of $r$ - the semi-major axis of the planet's orbit in units of $R_*$ - the stellar radius.
Following \cite{H2011}, we count each planet with a semi-major axis $r$ as $r$ planets, to account for planets in other systems in which the planets are not transiting due to unfavorable inclination \citep{BS1984}. Note that this factor should be applied to each transiting planet regardless of the planarity of the system and the correlations it creates.
In fig.\ref{fig:cumulative} we plot the calculated cumulative number of planets with radii smaller than $R_{Jupiter}/3$ as well as for planets with radii larger than $R_{Jupiter}/3$.
At a distance of $r_{\max}\sim 75$ the smaller planets experience a role over, whereas the larger planets do not (see figure \ref{fig:cumulative}).
We interpret this break as the distance at which the completeness of the sample, starts to break for smaller planets.
This interpretation is based on the assumption that the distribution of small planets and of large planets as a function of their semi major axis is similar as was found by \cite{You2011}.
This break gives the upper boundary of our sample which corresponds to a period of 75.5 days.

Since the exact location of the role over is unclear and may be dependent on the definition of small and large planets, one may argue that the uncertainty in the location of $r_{\max}$ may affect the applicability of our results.
Therefore, in \S \ref{sec:complete}, we discuss the applicability of our result in the case where the data is assumed to be complete up to distances smaller than $75R_*$.
We find the conclusions to be quantitatively similar for any $r_{\max}>45R_*$.
Therefore, from now on and up to \S \ref{sec:complete}, we assume the data is complete up to $r_{\max}=75$ and restrict our data analysis for distances smaller than that.

The probability of hosting a planet within the range $\{r,r+dr\}$ is given by $f(r)dr$ where
$f(r)$ is defined to be the occupancy distribution.

At the range $\{0,r_{\max}\}$, we fit $f(r)$to some piecewise power-low function of the form:
\begin{equation}\label{fr}
f(r)=
\left\{
\begin{array}{c c c}
0 & & r<r_a \\
\lambda\left(\frac{r}{r_b}\right)^2 &   &  r\in\{r_a, r_b\}\\
\lambda   &   & r\in\{r_b, r_c\}\\
\lambda\left(\frac{r}{r_c}\right)^{-1}   &  &  r\in\{r_c, r_{\max}\}\\
0 & & r>r_{\max} \\
\end{array}
\right.
\end{equation}
In order to calculate $r_{a},r_{b}$ and $r_{c}$, we minimize the result of the chi-square between the empirical cumulative distribution function and the theoretical function with regards to $r_{a},r_{b}$ and $r_{c}$.
We find: $r_a=3.4$, $r_b=12.8$ and $r_c=30.8$ for $r_{\max}=75$.

\begin{figure}
  % Requires \usepackage{graphicx}
  %\includegraphics[width=0.8\textwidth]{ecdf}\\
  \plotone{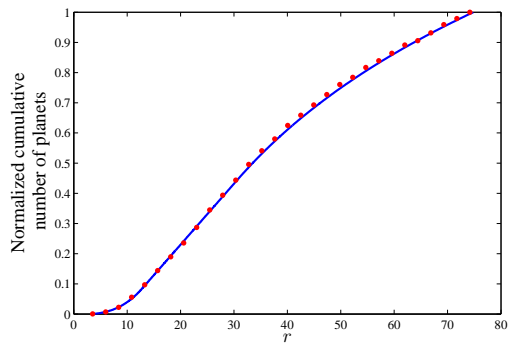}
  \caption{The normalized cumulative number of planets as a function of $a/R_*$. The blue solid line is the analytical fit for $\int f(r)dr$ while the red dots are from Kepler's data.}\label{fig:ecdf}
\end{figure}

Figure \ref{fig:ecdf} shows our fitted cumulative distribution function compared to the observed data.
It is evident that the functional form described by equation \ref{fr} is sufficient to adequately describe the data.

Note, that although there is a possibility that the current data may be supplemented by the discovery of additional, non transiting, planets, e.g. through TTV's \citep{TTVagol,TTVholman,F2011,F2012},
this would not affect our analysis since those were already taken into account statistically, by counting each transiting planet as $r$ planets.
In this paper, we compare the expected distribution of transiting planets with Kepler's data.

\section{The Independent Planar Model}\label{planar}
\subsection{Model Assumptions}

We propose a very simple model in attempt to explain the amount of systems with some number $m$ of transiting planets. This simple model allows us to evaluate these numbers analytically.

The basic assumptions of the independent planar model are as follows:
\begin{enumerate}
\item {\textbf{All planets in a system are exactly aligned} -
This assumption maximizes the correlation between different planets in a system and fortunately,
it allows us to analytically model the probabilities of having a given number of transiting planets in a given system.
This assumption only serves us in the analytic discussion, while later
we also discuss the case of partially misaligned planetary systems (\S3.2).
}

\item {\textbf{All of the stars and planets are identical} -
Taking only F, G and K stars from the Kepler data, we limit ourselves to stars which have comparable mass and radii.
By not including planets that are detected above a certain dimensionless semi-major axis $r$,
we effectively reduce the effect of the planet's size on the detection efficiency.}

\item {\textbf{The Occupancy distribution} of a planet existing at a given distance from its stellar host, $f(r)$, is the same for all the stars which are capable of producing planets
and is given by equation (\ref{fr}).}
This assumption implies that in a given stellar system occurrence of a planet at one position is independent of the occurrence of other planets at other distances.
\end{enumerate}
The basic assumptions of the independent planar model are similar to some aspects of models previously suggested \citep{Lis2011,TD2011,You2011}.
A detailed comparison between this model and its results to those of previous works, are discussed on \S \ref{sec:compare}.

\subsection{$P(m)$ - The probability for $m$ transiting planets}
We find $P(m)$, the probability that a given star hosts $m$ transiting planets,
for a general model, in which the occupation probability function has a general form of $f(r)$.
In the following we denote distances in units of $R_*$.

The zero mutual inclination assumption guaranties that for each orientation of the plane, there is a maximal distance from the star beyond which a planet does not transit.

The probability for $m$ transiting planets around a given star is therefore:
\begin{equation}\label{comb1}
\begin{split}
P(m)&=\int_0^{\infty} dr
 \left(
  \begin{array}{c}
   \mbox{Probability density that the maximal}\\
   \mbox{transit distance is }r .
  \end{array} \right)\\
&\times
\left(
  \begin{array}{c}
   \mbox{Probability that }m \mbox{ planets}\\
   \mbox {exist at distances smaller than }r.
  \end{array} \right).
\end{split}
\end{equation}

The probability that $m$ planets exist at distances smaller than $r$ can be found from
the mean number of planets up to distance $r$
\begin{equation}\label{Fndelta}
F(r) \equiv \int_{0}^{r}dr'\ f(r').
\end{equation}
Assuming no mutual influences between planets, the number of planets up to some radius is a Poisson variable.
Therefore, the probability that $m$ planets exist at distances smaller than $r$ is:
\begin{equation}\label{combPkm}
\left(
  \begin{array}{c}
   \mbox{Probability that }m \mbox{ planets}\\
   \mbox {exist at distances smaller than }r\\
  \end{array} \right)
=
\frac{F(r)^m}{m!} e^{-F(r)}\ .
\end{equation}

Given a random orientation of the plane of the system, the probability of an object located at radius $r$ to transit is $1/r$
\citep{BS1984}.
Therefore, the probability density that the maximal transiting distance is $r$ is given by $1/r^2$.
Substituting this result and equation (\ref{combPkm}) into equation (\ref{comb1}) we obtain:
\begin{equation}\label{comb2}
P(m)=\int_0^{\infty} dr
\left[
\frac{F(r)^m}{r^2m!} e^{-F(r)}
\right].
\end{equation}

In reality, there is a radius, $r_{\max}$, beyond which transiting planets are less likely to be
detected due to the finite duration of the observations.
Therefore, we limited our sample to $r<r_{\max}$ so that $f(r>r_{max})=0$.
Consequently, $F(r>r_{\max})=F(r_{\max})$, and
equation (\ref{comb2}) becomes:
\begin{equation}\label{comb3}
P(m)=\int_0^{r_{\max}} dr
\left[
\frac{F(r)^m}{r^2m!} e^{-F(r)}
\right]+
\frac{F(r_{\max})^m}{r_{\max}m!} e^{-F(r_{\max})}
\end{equation}

\subsection{$P(m)$ for our Occupancy Distribution}
So far, we have found $P(m)$ for a general form of $f(r)$.
At this point, we find $P(m)$ for our $f(r)$ given by
equation (\ref{fr}).

$P(m)$, is a function of $f$ via $F$. Therefore, we can decompose it into four different parts
\begin{equation}\label{P123}
P(m)=P_a(m)+P_b(m)+P_c(m)+P_d(m)\ ,
\end{equation}
where $P_a(m),P_b(m)$ and $P_c(m)$ correspond to the different regimes of $f(r)$,
and $P_d(m)$ is related to the last term in  equation (\ref{comb3}).
$F(r)$ is simply given by:
\begin{equation}\label{Fofr}
\begin{split}
F(r)&=  \int_0^r dr\ f(r) \\
&=\left\{
\begin{array}{c c c}
0 &  &   r\in\{0, r_a\}\\
A\left(r^{3} -  r_a^{3} \right)  &   &   r\in\{r_a, r_b\}\\
F(r_b)+\lambda (r-r_b)  &  &   r\in\{r_b, r_c\}\\
F(r_c) +  \lambda r_c \ln (r/r_c)   &   &   r\in\{r_c, r_{\max}\}\quad\\
\end{array}\right.
\end{split}
\end{equation}
where $A=\lambda/3r_b^2$.
This gives us:
\begin{subequations}\label{ver1}
\begin{align}
P_a(m)=  &
\int_{r_a}^{r_b}dr\ \left[ \frac{F^m}{r^2m!} \ e^{-F} \right]
\quad
\\
P_b(m)=  &
\int_{r_b}^{r_c}dr\ \left[ \frac{F^m}{r^2m!} \ e^{-F} \right]
\\
P_c(m)=  &
\int_{r_c}^{r_{\max}}dr\ \left[ \frac{F^m}{r^2m!} \ e^{-F} \right]
\\
P_d(m)=  &
\frac{F(r_{\max})^m}{r_{\max} m!}  e^{-F(R_{\max})}\quad .
\end{align}
\end{subequations}

%The planar assumption is not adequate in explaining Kepler's observations
Since the number of systems with zero planets is vastly
larger than these with the planets,
%underestimated in order to fit the number of systems with planets.
we add another free parameter, $C$, which determines the fraction of stars that are able to produce planets with efficiency $\lambda$. Note that we still require that stars that are capable of producing planets to be identical.
In this case:
\begin{equation}
\begin{split}
&P^C(m)= \\
&\left\{
\begin{array}{c c}
 C\left[P_a(m)+P_b(m)+P_c(m)+P_d(m)\right]  &   \text{for $m\ne 0$} \\
\\
 1-\sum_{m'=1}^\infty P^C(m')  &   \text{for $m = 0$} \\
\end{array}\right.\ .
\end{split}
\end{equation}
In a sample of $N_s$ stars, the expected number of systems that contain $m$ transiting planets is
\begin{equation}\label{Nexp}
N_{exp}(m)=N_sP^C(m)\ .
\end{equation}

\section{Comparison to observed data}\label{obs_data}
Now that we have an analytical independent planar model for $N_{exp}(m)$ as a function of $\lambda$ and $C$, we can compare this model to the observed data.

The number of systems, with a given number of transiting planets as observed by Kepler which satisfy our completeness constrains, as presented in \S \ref{od}, is shown in table \ref{table1}.
Note that the number of systems that do not contain any planets is hard to estimate due to the fact that not all of the stars were observed continuously \citep{K2012}.
We set this number by hand to be $1.5\cdot 10^5$, where the exact number may differ from this amount and be between $(1.28-1.9)\cdot 10^5$.
However, the exact number is degenerate with our parameter $C$ and has no influence on the parameter $\lambda$,  therefore it does not affect the applicability of our result.

\begin{table}
\caption{The number of systems with $m$ \textit{transiting} planets as observed by Kepler, $N_{obs}(m)$, as well as the number of systems with a given amount of \textit{transiting} planets as expected from the independent planar model, $N_{exp}(m)$, all under our constrains as presented in \S \ref{od}.}
\begin{tabular}{c  c  c}
  \hline\label{table1}
  $m$   & $N^{obs}(m)$ & $N^{exp}(m)$ \\ \hline
  0 & 150,000  & 150,000\\
  1 & 1,185 & 1137.8\\
  2 & 204 & 271.9\\
  3 & 69 & 62.1\\
  4 & 20 & 11.96\\
  5 & 8 & 1.94\\
  6 & 0 &  0.27\\
\hline
\end{tabular}
\end{table}

\subsection{Zero Mutual Inclination}\label{ZeroMI}
\begin{figure}
  \plotone{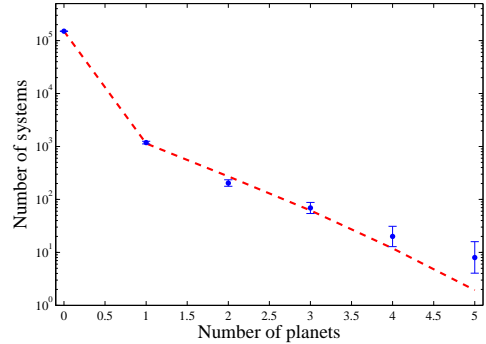}
  \caption{The number of systems with $m$ planets. Kepler's data is denoted by the blue dots while our analytical theory is plotted with the red dashed line. The black dashed line shows the best fit Poisson distribution for comparison. The error bars are the $95\%$ confidence level assuming each bin has a Poisson distribution.
 Note that the number of double systems is significantly underestimated, by more than $4\sigma$.}\label{fig:ana_fit}
\end{figure}

In order to find $\lambda$ and $C$ which best fit the
observations, denoted by  $\lambda_0$ and $C_0$  \footnote{The subscript "0" is for zero mutual inclination between the planets in the system.
Later on, when we discuss non-zero mutual inclination between the planets,
the subscript "0" will be replaced by $I$.},
we define a likelihood function $L_{obs}(\lambda, C)$, which represents the likelihood of having $\vec{N}_{obs}$ systems with $\vec{m}$ transiting planets, given an expectation value $\vec{N}_{exp}(\lambda, C)$
\begin{equation}\label{Like}
L_{obs}(\lambda,C)=\prod_{m=1}^{\infty}\left(
\frac{[ N_{exp}(m)]^{ N_{obs}(m)}}{ N_{obs}(m)!}e^{- N_{exp}(m)}
\right)\quad
\end{equation}
This definition arise due to the fact that $N_{exp}(m)\ll N_s$ implying that $N(m)$ is a Poisson variable with an expectation value $N_{exp}(m)$.
We find that $L_{obs}(\lambda,C)$ has a maximum when $\lambda_0=1.78\cdot 10^{-2}, C_0=0.37$, and its value is
$L_{obs}(\lambda_0,C_0)=1.88\cdot 10^{-15}$. Figure \ref{fig:ana_fit} shows the best fit model compared to the observations of Kepler along with the best fit Poisson distribution.
As expected, the planar model fits Kepler's data better than the Poisson distribution since planarity adds an additional correlation.
The data agrees with the fit for $m=1$ transiting planets, overestimate the number of pairs and underestimate the number of systems with $m\geq 3$ transiting planets (see also \cite{doubles} for similar results).
This can be understood as a partial failure of the independent planar model in its attempt to overcome the long tail problem that ruled out the Poisson distribution model presented before.
In other words, the independent planar model does create an inherently longer tail compared to the Poisson distribution, but the tail is not long enough.

In order to estimate whether or not this deviation of the data from the predictions of the model is significant,
we compare $L_{obs}(\lambda_0,C_0)$ to the typical likelihood; i.e the expectancy value of the likelihood of sets of large number of $\vec{N}(m)$ drawn from a Poisson distribution with expected value $\vec{N}_{exp}(m)$.

The typical likelihood can be calculated as follows.
The number of systems with $m$ transiting planets is a Poisson variable with a mean $N_{exp}(m)$.
The probability of detecting $N$ systems with $m$ transiting planets, is
$[N_{exp}(m)^{N}/N!]e^{- N_{exp}(m)}$,
therefore the contribution to the typical likelihood that comes from systems with $m$ transiting planets is
\begin{equation}
\begin{split}
&L^m(\lambda_0,C_0)=\\
&
\sum_{N=0}^{N_s}
\left(
\frac{[ N_{exp}(m)]^{ N}}{ N!}e^{- N_{exp}(m)}\cdot
\begin{array}{c}
\text{\footnotesize{probability that}}\\
\text{\footnotesize{$N$ systems}}\\
\text{\footnotesize{contain $m$ planets}}\\
\end{array}
\right)
\end{split}
\end{equation}
where $N_s$ is the total number of stars in the sample.
Since the probability that $N$ systems contain $m$ planets is, again,
$[N_{exp}(m)^{N}/N!]e^{- N_{exp}(m)}$, we obtain:
\begin{equation}
L_{typ}^m(\lambda_0,C_0)=
\sum_{N=0}^{N_s}
\left(
\frac{[ N_{exp}(m)]^{ N}}{ N!}e^{- N_{exp}(m)}
\right)^2.
%\right]
\end{equation}
Since the typical likelihood is given by the product of $L_{typ}^m(\lambda_0,C_0)$ for all possible $m$'s, we obtain:
\begin{equation}\label{Ltyp}
L_{typ}(\lambda_0,C_0)=\prod_{m=1}^{\infty}\left[
\sum_{N=0}^{N_s}
\left(
\frac{[ N_{exp}(m)]^{ N}}{ N!}e^{- N_{exp}(m)}
\right)^2
\right].
\end{equation}

We find
$L_{typ}(\lambda_0,C_0)=5.18\cdot 10^{-8}$.
We define the ratio $L_{obs}(\lambda_0,C_0)/L_{typ}(\lambda_0,C_0)$ to be the success rate of the model.
This parameter is a proxy to the validity of the model,
values around unity indicate that the model fits the data well, while small values indicate improbable models.
Since in our case the success rate is $3.64\cdot 10^{-8}$,
we find that the independent planar model which assumes a planar system with independent planet occurrence,
can be disqualified with high certainty.

\begin{figure}
 \centering
 \plotone{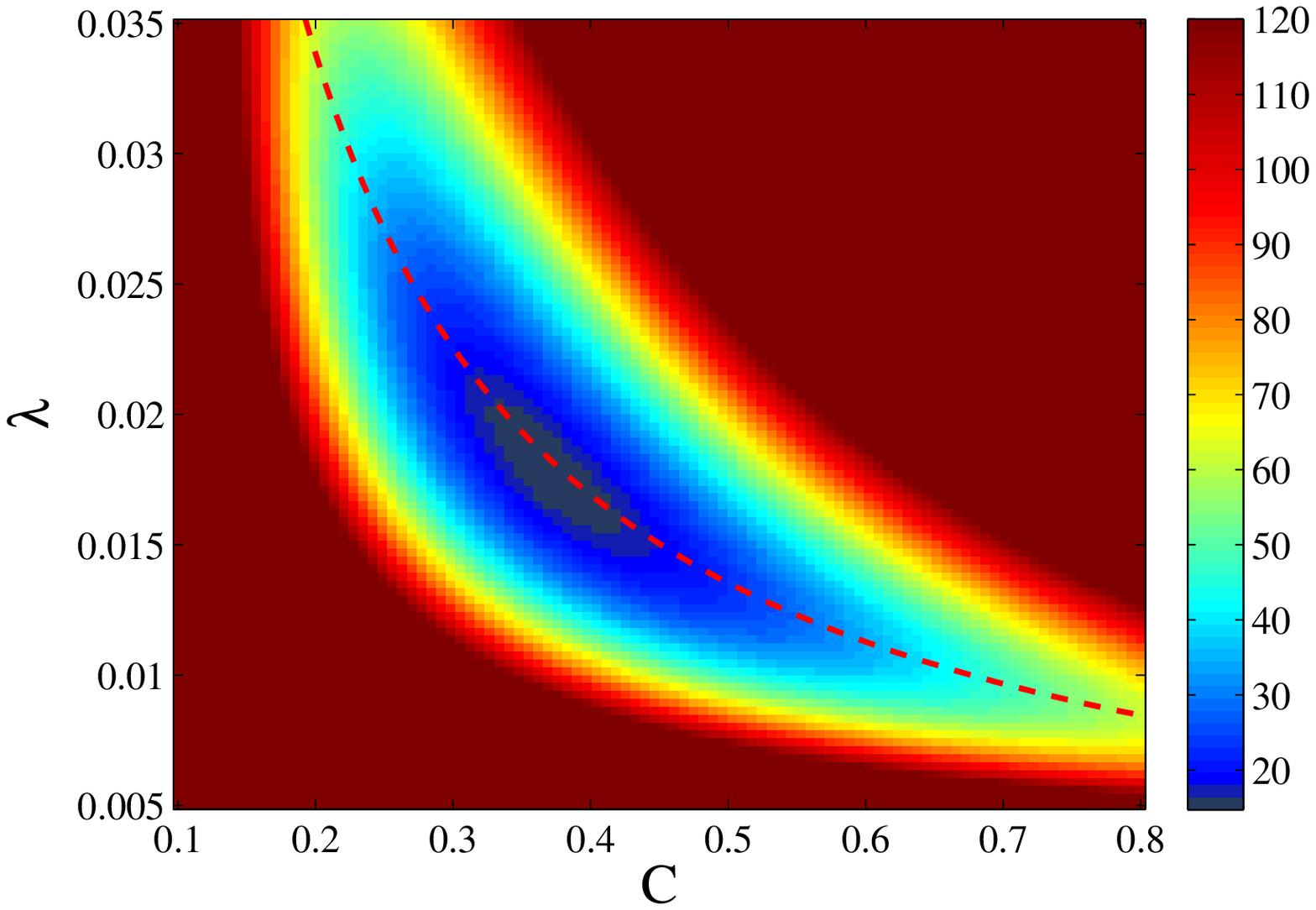}
   \caption{\footnotesize{$-\log_{10}(L_{obs})$ as a function of $\lambda$ and $C$ for zero mutual inclination.
The red dashed line shows $\lambda\propto C^{-1}$.\label{pretty_pic}
}}
\end{figure}
An interesting property of the likelihood function is that it depends strongly on the product $\lambda\cdot C$.
This product is proportional to both the total number of planets per star and to the number of transiting planets per star.
Specifically, for our function $f(r)$ given by equation (\ref{fr}), $46.5 C\lambda$ is the mean number of planets and $1.9 C\lambda$ is the mean number of transiting planets, regardless of the mutual inclination.
It is only the distribution of the number of transiting planets per system that depends on the mutual inclinations.
Along the curve $\lambda\propto 1/C$, the likelihood does not change significantly.
Figure \ref{pretty_pic} shows the value of the likelihood function in the $\lambda-C$ plane.

\subsection{Finite Mutual Inclination}\label{nonplanar}
In nature planetary systems are more likely to have a small mutual inclination between planets due to various processes\footnote{Processes that excite the eccentricity usually increase the mutual inclination between planets \cite[e.g.,]{review,C2008,J2008}.}. There is not much observational information about the mutual inclinations between planets. \cite{H2010} and \cite{Lis2011} placed an upper limit of $I<10^o$ on the mutual inclination of the systems Kepler-9 and Kepler-11, and \cite{B2011} found $I=5^o-20^o$ for the Kepler 10 system.
However, no statistical analysis has been able to rule out the possibility that the typical mutual inclination may be larger.

We now extend our treatment to include systems with mutual inclinations between planets. An analytical derivation for the general case is much more complicated than the special case with zero mutual inclination; the results in this section are purely numerical.

We expect that by accounting for finite mutual inclination the model of independent planets would deviate even more from observations, since it will reduce the dependency in the detection process which will lead to
fewer systems with a large number of transiting planets.

The mutual inclination is incorporated by having each planet randomly change its inclination relative to the system's plane. The random inclination is drawn from a Rayleigh distribution with a mode $I$.
We inspect systems with $I\in\{2^o, 5^o,10^o,20^o\}$, this adds an additional free parameter to the independent planar model.

In order to evaluate $P^C(m)$, the probability for $m$ transiting planets around a given star, in the case of
non zero mutual inclination we use a numerical simulation.
In each run of the simulation, we produce a planetary system that generates planets with a probability $C$.
If the system is capable of producing planets, we construct $\sim 4300$ bins ranging from $r_a$ to $r_{\max}$ in such a way that the probability for each bin to host a planet is the same and in accordance with $f(r)$ as presented in equation (\ref{fr}). Each bin is then assigned a random number that determines if it hosts a planet.

We then randomly draw an inclination for the whole planetary system from a uniform distribution of
$-1\leq \cos(i_s)\leq 1$ where $i_s$ is the system's inclination relative to the observer.
For each planet we randomly draw an orbital plane whose inclination relative to the plane of the planetary system, $I$, is drawn from a Rayleigh distribution and its argument of the ascending node is drawn from a uniform distribution. The inclination between the planet's orbital plane and the observer, $i_p$, is then computed.

Counting the number of planets satisfying $|r\cos i_p|<1$, we calculate how many planets transit
in that specific system. This process is then reproduced for the number of stars in the sample $N_s\simeq 1.5\cdot 10^5$, and by averaging this simulation on $10^3$ runs we find numerically $N^I_{exp}(m)$, the expected number of systems with $m$ transiting planets, under the assumption of a typical mutual inclination $I$.

After finding $N_{exp}^I(m)$ we find the maximum likelihood of the observations using equation (\ref{Like}) where $N_{exp}(m)$ is replaced by $N_{exp}^I(m)$ and denote the $\lambda$ and $C$ that correspond to this maximum to be $\lambda_I$ and $C_I$.

As was done in the previous section, we compare $L_{obs}^I(\lambda_I, C_I)$ to the typical likelihood expected for $I,\ \lambda_I$ and $C_I$ (using equation (\ref{Ltyp}) where  $N_{exp}(m)$ is replaced by $N_{exp}^I(m)$).
As done in \S\ref{ZeroMI},
we define the ratio $L_{obs}^I(\lambda_I, C_I)/L_{typ}^I(\lambda_I, C_I)$ to be the model's success rate.

The opposite extreme to the independent planar model is
the case of an isotropic distribution of planets. It can be described by a simple analytical expression.
If planets have an isotropic distribution, then each cell has some small chance (but independent from the other cells) of hosting a transiting planet. This naturally gives rise to a Poisson distribution. However, a Poisson distribution by itself hardly fits the observational data \citep{Lis2011}.
The analog of an isotropic distribution to the discussed model is having a fraction, $C$, of stars being capable of hosting planets. Those that are capable of hosting planets have a Poisson distribution of the number of transiting planets with a mean $\mu$. The probability of a system hosting $m$ transiting planets is then
\begin{equation}
\left.
\begin{array}{c c}
P^C(m)  = C\frac{\mu^{m} e^{-\mu}}{m!} &\text{for}\ m\neq 0 \\
  &   \\
\begin{split}
P^C(0)  & = 1-\sum_{m'\neq 0}P^C(m')\\
& = (1-C)+Ce^{-\mu} \end{split}
&\text{for}\ m= 0 .
 \\
\end{array}\right.\label{Poisson}
\end{equation}
The Poisson parameter $\mu$ is related to $\lambda_{Isotropic}$ by:
\begin{equation}
\mu=\int_0^{\infty}dr \frac{f(r)}{r}
\end{equation}

Table \ref{inclination} contains $\lambda_I$, $C_I$ and the success rate for the cases of
$I\in\{2^o,\ 5^o,\ 10^o,\ 20^o\}$ as was found from the numerical simulations.
In addition, the table also contains the parameters for the planar case, as was found in \S\ref{ZeroMI}, as well as for the isotropic case.
\begin{table}
\begin{tabular}{ c c c c c}
Inclination & $\lambda_I$ & $C_I$ &   Success Rate \\
\hline
$0^o$\ &    $1.78\cdot 10^{-2}$ &     $3.7\cdot 10^{-1}$ &     $ 3.64\cdot 10^{-8}$  \\
$2^o$ &   $(2.3\pm0.1)\cdot 10^{-2}$ &     $(2.84\pm0.05)\cdot 10^{-1}$ &        $ 4.1\cdot 10^{-10}$ \\
$5^o$ &   $(3.7\pm0.11)\cdot 10^{-2}$ &     $(1.78\pm0.05)\cdot 10^{-1}$ &      $ 1.51\cdot 10^{-11}$  \\
$10^o$ &   $(6.5\pm0.22)\cdot 10^{-2}$ &     $(9.8\pm0.3)\cdot 10^{-2}$ &       $ 3.94\cdot 10^{-12}$ \\
$20^o$ &   $(1.24\pm 0.04)\cdot 10^{-1}$ &     $(5.3\pm0.16)\cdot 10^{-2}$ &       $ 3.39\cdot 10^{-12}$ \\
Isotropic &    $2.78\cdot 10^{-1}$ &     $2.36\cdot 10^{-2}$ &        $ 2.93\cdot 10^{-18}$ \\
\hline
\end{tabular}\caption{Summary of our best fit results for different mutual inclinations.
$\lambda_I$ is proportional to the probability of having a planet in a given cell,
$C_I$ is the fraction of stars that are capable of hosting planets and
the success rate is the ratio $L_{obs}^I(\lambda_I, C_I)/L_{typ}^I(\lambda_I, C_I)$ which represent the probability that the independent model fits the observations.
}\label{inclination}
\end{table}

\subsection{Conclusions from the comparison to the observations}

As one can see from table \ref{inclination}, all possible inclinations are very unlikely (success rate$\ll 1$).
Therefore, our main conclusion is that Kepler's observations can not be explained by a model which assumes that planet occurrence is an independent process
and all planet producing stars are identical.
Some additional correlation is required.

In the next section, we discuss briefly the types of correlations that might explain Kepler's data.

\section{Possible Explanations}
\label{sec:depend}
The inapplicability of the independent models presented before,
might arise due to differences in the star's planet producing efficiency, and/or
if planets in a given solar system affect each other, either in terms of creation or dynamical evolution.
In this section, we present two possible explanations that may cause the failure of the independent model,
one which arises from dependency between planets in a given solar system
and one which arises from differences in the star's planet producing efficiency.

\subsection{Mutual Planet Dependence}\label{Mutual_Planet_Dependence}

Previous works have shown that planets in a given stellar system tend to be relatively close to resonances \citep{Lis2011,Lis2012res}. \cite{Latham} have shown that systems that contain giant planets (in particular hot Jupiters) tend to be single systems rather than multi.
This indicates that mutual planet dependence might play a role, i.e. planets in a given stellar system affect the semi major axes of one another.
In this section, we investigate this correlation.

Assuming that planets occurrence is an independent process and taking into account only pairs of transiting planets,
we can calculate the distribution of the ratio $r_2/r_1$, where $r_2$ and $r_1$ are the semi major axes of the distant and the close planets respectively.
This distribution, which depends on $f(r)$, can be found numerically and compared to the distribution of $r_2/r_1$ from the pairs in Kepler's data.

\begin{figure}
 \centering
 \includegraphics[trim = 00mm 00mm 00mm 00mm, clip, width=8cm]{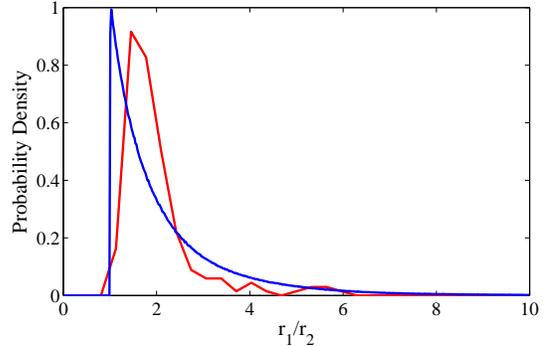}
   \caption{\footnotesize{
The distribution of the ratio $r_2/r_1$.
The red line represents the probability density of Kepler's $204$ pairs and the blue line represents the probability density for our synthetic pairs.
}}
\label{ratio_in_pairs_pic}
\end{figure}

In figure \ref{ratio_in_pairs_pic}, we present the distribution of the ratio $r_2/r_1$ from Kepler's $204$ pairs (red line) compared to the distribution of the ratio $r_2/r_1$ from our synthetic pairs (blue line).

The distribution from Kepler's data has a deficit compared to the synthetic pairs both in the similar radius pairs as well as the distant pairs.
Using a two sample KS test, we find that the two distributions can not arise from the same distribution (the null hypothesis is rejected at level $\simeq 10^{-10}$).
This conclusion rejects any model which assumes that the location of planets in a given system is independent on the existence of other planets in the system.

The nature of the dependency in a given system can be describe qualitatively.
We know that planets can not be too close to each other due to consideration of dynamical stability \citep[e.g.]{Lis2011}, this explains the deficit in pairs with the same radius.
On the other hand, it is evident that planets in a given system tend to be closer to one another than expected from an independent distribution, this is also demonstrated by their tendency to be near low-order mean-motion resonances \citep{Lis2011,Lis2012res}, this explains the deficit of planets with a large radius ratio.

Note that a dependency in the location of different planets in a given solar system naturally affects the probability of planet detection.
The expected number of systems with some specific number of planets $m$, is influenced by this dependency.
Therefore, we find this dependency to be at least a partial cause to the observed deviations from the independent models expectations.

\subsection{Fertile and Sterile Systems}

Some previous works have argued that metal rich stars are more likely to produce planets \citep{FV2005,metal1,metal2,JL2012,metalNature}.
It is not unreasonable to assume that the normalization or even the functional form of $f(r)$ may depend on the metallicity of the star.

In order to see whether or not a metallicity dependence of $\lambda$ is sufficient to produce the observed correlation,
we divide the planets producing stars into two different populations characterized by two different values of $\lambda$.
Note, that this model is not an exact model, but rather an order of magnitude calculation done in order to see weather metallicity's dependence as the one presented by previous works, may explain Kepler's results.

The planet producing efficiency seems to increase by a significant factor
at metallicity $[Fe/H]=0.1$ compared to $[Fe/H]=-0.18$ \citep{FV2005}.
According to \cite{PR2012}, the metallicity distribution among Kepler's stars is $[Fe/H]=-0.18\pm 0.28$.
Guided by these results, we divide the planets producing stars into two categories:
one, which contain $80\%-90\%$ of the planets producing stars and characterized by planet producing efficiency $\lambda$, and the other, contain a fraction of $F_Q\sim 10\%-20\%$ of the planet producing stars, which
characterized by planet producing efficiency $Q\lambda$, where $Q$ is in the range $3-10$.
Maximizing the likelihood function (equation \ref{Like}),
we find a good fit to the observations ($L_{obs}\sim 10^{-1}L_{typ}$) both for an independent planar model as well as for isotropic model
(see table \ref{tab:Two_Lambdas}).

\begin{table}
\begin{tabular}{ c c c c c c}
   & $\lambda$ & $Q$ &$C$ &  $F_Q$ & $L_{obs}/L_{typ}$ \\
\hline
Planar &  $4.3\cdot 10^{-3}$  &   $6.7$  &  $7.2\cdot 10^{-1}$  &   $20\%$  &   0.20  \\
Isotropic &  5.2$\cdot 10^{-2}$  &   $8.1$  &   $7.4\cdot 10^{-2}$  &   $10\%$  &   0.11  \\
\hline
\end{tabular}\caption{
Summary of our best fit results allowing two different planet producing efficiencies. $Q$ is the ratio of planet producing efficiency and $F_Q$ is the fractional portion of the high efficiency population.
}\label{tab:Two_Lambdas}
\end{table}

It is therefore conceivable that differences in the metallicities between different stars may be responsible for the correlations we observed in the data.

Note that the correlation between metallicity and planet producing efficiency is not unambiguous and resent works have shown different results.
\cite{metal2} and \cite{metalNature}, show that this correlation only holds for Neptune sized planets and above, or for M dwarf stars. Since Kepler's data contains mostly smaller planets where the metallicity has a weaker influence on the star's planet producing efficiency,
another mechanism with such significant influence on the $\lambda$ distribution, may explain the deviations from the independent model.

\section{Comparison With Previous Works}
\label{sec:compare}
\cite{Lis2011} used the data from Kepler's first data release to estimate statistically the coplanarity of planetary systems.
Their sample included all planets orbiting stars with $R_*<10R_\odot$, with periods $3<P<125$ days, radii $1.5<R_p<6$ Earth radii and $\textrm{S/N}>16$ (as listed for Q0-5
\footnote{Q0-5 symbolizes data taken from the first six quarters of Kepler.}
). They numerically create a synthetic population of planets by assigning each star in the Kepler's target list a number of planets drawn from either a Poisson, uniform or exponential distribution. The planets are then given periods and sizes to match the observed population while maintaining a minimum radial separation between planets in order to preserve the system's stability. Each planetary system is then randomly oriented and each planet is given a random inclination. The simulated planets are then checked to see if they would have been detected by Kepler. They find that a low-mutual inclination gives the best result and that only $3\%-6\%$ of stars host planetary systems. However, they can not rule out the possibility that Kepler's data contains stars with a large amount of planets and a high mutual inclination. This work is most similar to our approach, especially the case where they choose the number of planets is chosen from a Poisson distribution, which is suitable for independent planets.
The reason that we arrive at different conclusions it that we use a larger data set than \cite{Lis2011}. We have also implement an analytical model rather than a numerical one which simplified the analysis. \cite{Lis2011} and us differ in our approach towards addressing the issue of data completeness. We empirically find the distance at which small planets are no longer detected and while \cite{Lis2011} mimic the stellar and planet population of Kepler and simulate the transits of a given planet.

Using the same stellar selection criteria as we do, \cite{TD2011} analyzed planets from Kepler's first data release with radii smaller than 2 Jupiter radius and periods up to 200 days. Taking into account geometric corrections, they fit a "planetary distribution function", a distribution function that specifies the probability of a star to host a given number of planets, to the observed data.
This approach is in some sense, the opposite extreme to ours. They allow for any planet number distribution, while our approach which assumes identical stars and independent planet results in a Poisson distribution of the number of planets.
It is for this reason that they find that the data can be explained by a small number of stars hosting a large amount of high mutual inclination planets as well as a large amount of stars hosting a smaller amount of planets in a rather planar configuration.
Like this work, they assume independence for the position of planets in a given stellar systems; i.e. the occurrence of one planet did not influence the semi major axes of the others.

However, in \S\ref{Mutual_Planet_Dependence} we have shown that there is a dependency between planets in a given planetary system. This finding seems to rule out the independent position assumption used both by \cite{TD2011}'s model as well as our simple independent planet model.

\cite{You2011} has tried to find the form of the "underlying planetary distribution function".
This function is somewhat similar to our $f(r)$, but it also contain a dependency on the radii of the different planets; i.e. it describes the probability of hosting a planet with a given radius, $R_p$, at some period $P$.
In order to find this function, \cite{You2011} only needed information regarding the total number of planets.
Therefore, he has used the total number of planets under a detailed analysis of survey selection effects. Our work tries to answer a different question. We assume the simplest underlying planetary distribution function that matches the observed period distribution and check to see if it is consistent with the observed multiplicities in the Kepler's data. This allows us to rule out the possibility that planet occurrence is independent of other planets.

\section{Summary and Conclusions}\label{summary}
It is clear that transiting planet occurrence cannot be a totally independent process;
transiting planets do not occur randomly in systems, since independent process
gives rise to a Poisson distribution in the number of systems with $m$ transiting planets, rather additional dependencies are required.

The simplest dependency one may assume is a perfectly planar solar system.
In this paper, we found analytically the probability that a given star hosts $m$ transiting planets, $P(m)$, assuming that planet occurrence is an independent process and allowing for planar systems.
The probability, $P(m)$, in this model, depends only on the parameter $\lambda$, which represents the planet producing efficiency of all stellar systems.
The planar assumption was not adequate in explaining Kepler's observations, so we added another free parameter $C$, which determines the fraction of stars that are able to produce planets. This model also failed to explained the observed data.

We conclude therefore that Kepler's observations can not be explained by any model which assumes that planet occurrence is an independent process with the same efficiency $\lambda$ for all planet producing stars. Some additional correlation is required.
We discuss two possible correlations that might explain the deviations from the simple independent model.

The first possible correlation is some mutual influence of planets in a given stellar system.
In attempt to find evidence for such mutual influence, we compared the ratio between semi-major axes in Kepler's pairs, to that of a synthetic population which were numerically produced under the assumption of no mutual influence.
We find that Kepler's data does not match the synthetic population.
Although this might not be the only deviation from the independent model, we speculate that this mutual influence is significant and it has an influence on the observed number of multiple planets.

Another possible correlation is that $\lambda$ is not constant among the different stars; i.e. planet producing stars are not identical and may produce planets with different efficiencies.
Using previous studies about the metallicity influence on planet producing efficiency, we have shown that taking into account a metallicity distribution might explain the deviation of the data from the presented model.
However, since the influence of metallicity seems to hold only for giant planets, this additional dependency should be taken as an example of how an additional dependency can solve the problem rather than be interpreted as the real physical dependency.

\begin{acknowledgments}
This research was partially supported by ERC and IRG grants and a Packard and Guggenheim
Fellowships. E.S. is partially supported by an Ilan Ramon grant from the Israeli Ministry of Science.
\end{acknowledgments}

\section{Appendix: Discussion about Completeness}
\label{sec:complete}

In section \S\ref{od}, we have discussed the importance of data completeness for statistical analysis as the one presented in this work.
In our work, finding $r_{\max}$, the maximal semi major axis for which the data is assumed to be complete,
was based on locating the point in which the number of small planets experience a role over.
Since the definitions of "small planets" and "role over" may be a bit flexible, we investigate the validity of our results for $r_{\max}$ which are different from $75R_*$.

\begin{figure}
 \centering
 \includegraphics[trim = 00mm 00mm 00mm 00mm, clip, width=8cm]{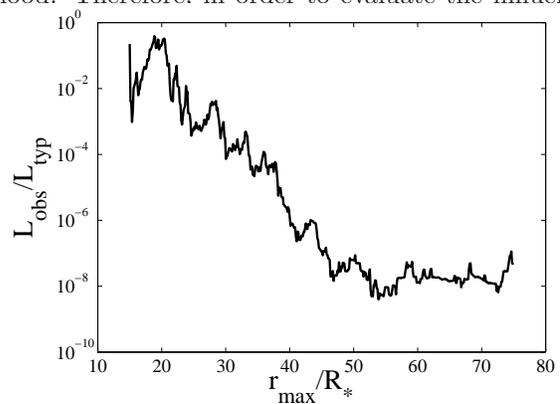}
 \caption{\footnotesize{
The success rate, i.e the likelihood of the best fit independent planar model compared to the typical likelihood as a function of $r_{\max}$, the maximal semi major axis for which the data is assumed to be complete.
\label{Lik_fu_r_max_pic}
}}
\end{figure}

The validity of the model is given by the successes rate, i.e. the maximal likelihood of the observations when $\lambda$ and $C$ are free parameters, compared to the typical likelihood.
Therefore, in order to evaluate the influence of the assumed $r_{\max}$ on the model,
we find the success rate of the model for every $r_{\max}\in\{15R_*,75R_*\}$ using the same procedure presented in \S\ref{ZeroMI}.
The success rate of the independent planar model is presented on figure \ref{Lik_fu_r_max_pic}\footnote{This procedure is done only the independent planar model.}.

From figure \ref{Lik_fu_r_max_pic}, one can deduce that for $r_{\max}> 25R_*$, the success rate is smaller than $10^{-2}$ and for $r_{\max}> 45R_*$ it is smaller than $10^{-7}$.
Therefore, changing $r_{\max}$ from the our canonical value of $75R_*$ by a factor of order unity, seems to have a small influence on our qualitative results.

\bibliography{reem}

\end{document}